\documentclass[useAMS,usenatbib]{mn2e}
\usepackage{amsmath,astrojournals,fleqn,graphicx,times}
\usepackage{natbibmnfix}

\title[Tailoring triaxial $N$-body models via a novel made-to-measure method]
      {\boldmath Tailoring triaxial $N$-body models via a novel
        made-to-measure method}

\author[Walter~Dehnen]
{
  Walter Dehnen%
\thanks{Email: walter.dehnen@astro.le.ac.uk}\\
  Department of Physics \& Astronomy,
  University of Leicester,
  Leicester, LE1~7RH
}

\date{Accepted .
      Received ;
      }
 
\pagerange{\pageref{firstpage}--\pageref{lastpage}}
\pubyear{2008}
\newcommand{\B}[1] {\boldsymbol{#1}}
 
\begin{document}

\maketitle
\label{firstpage}

\begin{abstract}
  The made-to-measure $N$-body method \citep{SyerTremaine1996} slowly adapts
  the particle weights of an $N$-body model, whilst integrating the
  trajectories in an assumed static potential, until some constraints are
  satisfied, such as optimal fits to observational data. I propose a novel
  technique for this adaption procedure, which overcomes several limitations
  and shortcomings of the original method. The capability of the new technique
  is demonstrated by generating realistic $N$-body equilibrium models for
  dark-matter haloes with prescribed density profile, triaxial shape, and
  slowly outwardly growing radial velocity anisotropy.
\end{abstract}
\begin{keywords}
  stellar dynamics -- 
  methods: $N$-body simulations -- 
  galaxies: kinematics and dynamics --
  galaxies: structure --
  galaxies: haloes
\end{keywords}

%%%%%%%%%%%%%%%%%%%%%%%%%%%%%%%%%%%%%%%%%%%%%%%%%%%%%%%%%%%%%%%%%%%%%%%%%%%%%%%%
\section{Introduction}
\label{sec:intro}

A standard problem in contemporary galaxy dynamics is the interpretation of
kinematic observations of galaxies in terms of their orbital structure as well
as their dark and luminous matter distribution. There are several methods one
can employ for this problem. First, moment-based methods find solutions of the
Jeans equations (or higher-order velocity moments of the collisionless Boltzmann
equation) that best fit the observed moments, such as density and velocity
dispersion. Secondly, distribution-function-based methods directly fit the
distribution function to the data, which can be more general than mere moments,
e.g.\ the line-of-sight velocities of many individual objects. Both techniques
are usually restricted to spherical or, under certain simplifying assumptions,
axisymmetric systems (though for different reasons\footnote{For moment-based
  models, symmetry reduces the number of independent moments and enables simple
  assumptions necessary to close the Jeans equations. For non-spherical
  distribution-function models, knowledge of isolating integrals of motion other
  than energy is necessary, and angular momentum is available only for axial
  symmetry.}).  However, distribution-function models are technically much more
challenging (since an integral equation has to be solved instead of differential
equations) and hence much less used than the moment-based approach.  In both
cases, astrophysically unjustified assumptions, such as velocity isotropy, are
often made in order to make the problem tractable.

Thirdly, Schwarzschild's (\citeyear{Schwarzschild1979},
\citeyear{Schwarzschild1993}) orbit-based method constructs a dynamical model by
first integrating many orbits over many orbital times in an assumed
gravitational potential, whereby recording their properties in an orbit library,
and then superposing them such that a best fit to the data is obtained. This is
a powerful method, since it comes, in principle, without restrictions on the
symmetry, and one may even obtain the distribution function
\citep{HafnerEtal2000}. However, in practice most applications are restricted to
axisymmetry, since there are several technical subtleties to overcome when
applying the method to potentials with a complex phase-space structure, the
typical situation for triaxial or barred systems \citep[though this is not
impossible and has been done, e.g.][]{HafnerEtal2000}.

Fourthly, in 1996, \citeauthor{SyerTremaine1996} (hereafter ST96) introduced
the `made-to-measure $N$-body method', which slowly adapts a $N$-body model to
fit the data. The first application of this method came as late as 2004, when
\citeauthor*{BissantzDebattistaGerhard2004} used it to construct a dynamical
model for the Milky Way's barred bulge and inner disk. More recently,
\citeauthor{DeLorenziEtal2007} (\citeyear{DeLorenziEtal2007}, hereafter DL07)
refined the method to incorporate observational errors; this has since been
applied for modelling elliptical galaxies to asses their dark-matter content
\citep{DeLorenziEtal2008a,DeLorenziEtal2008b}. The made-to-measure (hereafter
`M2M') method is as powerful as Schwarzschild's orbit-based method, and in
fact is closely related. Whereas in Schwarzschild's approach orbits are first
separately integrated and then superimposed, these two steps are merged in the
M2M method: trajectories are integrated and their weights adapted at the same
time. As a consequence there is no need for an orbit library and all the
technical difficulties associated with it. However, with the M2M method as
proposed by ST96 \& DL07 some problems remain, as I shall discuss, in
particular the appropriate time scale for adapting the particle weights of the
$N$-body model.

Finally, \cite*{RodionovAthanassoulaSotnikova2009} introduced a variation of
the M2M technique (though the authors did not make this association), which
they dubbed the `iterative method'. Their method starts from a
near-equilibrium dynamical model (constructed by any other method), which is
alternately relaxed under self-gravity (to evolve towards equilibrium) and
adapted to prescribed properties. In practice, this method too employs the
$N$-body approach and, like the traditional M2M technique, suffers from the
time-scale problem.

Another application of all the aforementioned techniques is the generation of
$N$-body initial conditions representing a galaxy or galaxy component (though
moment-based models additionally require the incorrect assumption of a
Gaussian velocity distribution and should not be used for this
purpose). However as mentioned above, distribution-function models, which are
the most commonly employed technique for generating $N$-body galaxy models,
are restricted to spherical (or, under simplifying assumptions, axial)
symmetry, which seriously limits their realism. Since the M2M technique works
directly with $N$-body data, it offers a simple and natural way to generate
$N$-body initial conditions with prescribed properties (ST96), in particular
non-spherical shape and non-isotropic velocity structure. As we shall see,
however, this requires some modification to the traditional M2M technique.

In this paper, I revisit the M2M method and propose several modifications
aimed at improving it, in particular in view of its application for tailoring
$N$-body initial conditions. I present the traditional ST96 \& DL07 version of
M2M in section \ref{sec:trad} and my modifications to the method in
\S\ref{sec:novel}, while \S\ref{sec:ics} presents some tests of tailoring
non-spherical and/or velocity-anisotropic $N$-body models.  Finally
\S\ref{sec:disc} discusses the results and concludes.

\section{Traditional M2M}
\label{sec:trad}
In this section, the M2M method as laid out by DL07 (which in turn was based on
ST96) is outlined, though with slightly different notation and conventions.

%\subsection{The merit function}
The fitting of the $N$-body model to the data is expressed as maximisation
problem: the $N$-body model shall maximise the \emph{merit function}
\begin{equation} \label{eq:merit}
  Q = \mu S - \tfrac{1}{2} C.
\end{equation}
Here, $C$ is the \emph{constraint function}, which measures the goodness of
fit of the $N$-body model to some target.  There are many possible choices for
$C$, but for now let us follow DL07 and consider a $\chi^2$-like measure of
the deviation of moments of the $N$-body model from target values
\begin{equation}\label{eq:C}
  C = \sum_{j=1}^n \left(\frac{Y_j-y_j}
    {\sigma_j}\right)^2.
\end{equation}
Here,
\begin{equation} \label{eq:y}
  y_j = \sum_i w_i\,K_j(\B{x}_i,\B{\upsilon}_i)
\end{equation}
are some \emph{moments} of the model defined via the \emph{kernel}
$K_j(\B{x}_i,\B{\upsilon}_i)$ and the particle weights $w_i\equiv
m_i/M_{\mathrm{tot}}$, while $Y_j$ are the \emph{targets} for those moments
and represent the observed data\footnote{This standard practice restricts the
  data to be just moments of the distribution function, and excludes, for
  example, the line-of-sight mean and dispersion velocity, which are functions
  of moments.  However, this restriction is not fundamental and the method can
  easily be extended to fit any function of moments, see \S\ref{sec:ics:aniso}
  for an example} with uncertainties $\sigma_j$.

Simply minimising $C$ is not a well defined procedure for two reasons: first,
there is no point in reducing $C$ well below the expectation value even if this
were possible (this would amount to `fitting the noise'); second, minimising $C$
may not be uniquely constraining the $N$-body model: there are, for instance,
many possible equilibrium models with the same density. Thus, in order to yield
a well-defined problem, one has to \emph{regularise} the merit function by a
penalty functional $S$ times a Lagrange multiplier $\mu$, which controls the
amount of regularisation. The penalty function is traditionally taken to be the
pseudo-entropy
\begin{equation} \label{eq:entropy}
  S = - \sum_i w_i^\star \log \frac{w_i^\star}{\hat{w}_i}
\end{equation}
with $w_i^\star\equiv w_i/\sum_jw_j$ the normalised weights and
$\{\hat{w}_i\}$ a pre-determined set of normalised weights, the so-called
priors. For general priors, $S$ defined in this way is the
\cite{KullbackLeibler1951} information distance (also known as `K-L
divergence') of the model corresponding to $w_i=\hat{w}_i$ from the actual
$N$-body model, i.e.\ $S$ penalises against deviations of the normalised
weights from the priors. Only for $\hat{w}_i\propto f_i^{-1}$, where $f_i$
denotes the value of the equilibrium distribution function corresponding to
$w_i=1/N$, does $S$ reduce to the true entropy of the $N$-body model (plus a
constant; this corrects statements made by ST96 \& DL07).

%\subsection{The force of change}
The idea of the M2M method is now to adjust the weights slowly such that $Q$ is
maximised. The standard method is to evolve the weights according to
\begin{equation} \label{eq:dotw} 
  \dot{w}_i = \epsilon\,w_i\,U_i
\end{equation}
with some \emph{rate of change} $\epsilon$ and the \emph{velocity of change}%
\footnote{Unfortunately, ST96 dubbed $\epsilon w_iU_i$ the `force of change',
  which is an inaccurate analogy since it is proportional to the first time
  derivative of the dependent variable. Below I introduce a method which
  indeed uses the second time derivate, for which the expression `force of
  change' is much more appropriate.}
\begin{equation} \label{eq:foc} 
  U_i = \frac{\partial Q}{\partial w_i}.
\end{equation}
For the particular choice (\ref{eq:C}) of the constraint function, this gives
\begin{equation} \label{eq:foc:trad} 
  U_i = \mu \frac{\partial S}{\partial w_i} -
  \sum_{j=1}^n \frac{Y_j-y_j}{\sigma_j^2}\,K_j(\B{x}_i,\B{\upsilon}_i).
\end{equation}
For sufficiently small $\epsilon$, integrating (\ref{eq:dotw}) will increase
$Q$ and eventually result in a $N$-body model for which $Q$ is maximal and the
$w_i$ no longer change. This method is similar to (and was in fact inspired
by) \cite{Richardson1972} - \cite{Lucy1974} iteration, though with a much
reduced step size.

%\subsection{Time-averaging} \label{sec:trad:ave}
Unfortunately, it is not as simple as that, because the merit function, being a
function of the randomly sampled particle trajectories, is itself a random
variable and fluctuates even with fixed weights. In order to suppress these
fluctuations, traditional M2M replaces the model moments $y_j$ in (\ref{eq:C})
with their time-averaged values $\bar{y}_j$, which are obtained by integrating
the differential equation
\begin{equation} \label{eq:mom:ave}
  \dot{\bar{y}}_j = \alpha (y_j - \bar{y}_j)
\end{equation}
starting with\footnote{Corresponding to
  $\bar{y}_j(t)=y_j(0)\,\mathrm{e}^{-\alpha t} + \alpha \int_0^t
  \mathrm{e}^{\alpha(t^\prime-t)} y_j(t^\prime)\,\mathrm{d}t^\prime$; ST96 \&
  DL07 give $\bar{y}=\alpha \int_{-\infty}^t \mathrm{e}^{\alpha(t^\prime-t)}
  y_j(t^\prime)\,\mathrm{d}t^\prime$ which results from integrating
  (\ref{eq:mom:ave}) from $t=-\infty$, a practical impossibility.}
$\bar{y}_j=y_j$ at $t=0$. If fitting to observed data with finite
uncertainties $\sigma_j$, this method has the virtue that the model
uncertainties due to $N$-body shot noise (which have been ignored in the
definition of the constraint function) are much reduced.

A problem with this time-averaging is that the computation of the derivatives
required for the velocity of change (\ref{eq:foc}) is no longer straightforward.
In fact, $\partial \bar{y}_j/\partial w_i$ is simply the time-averaged kernel
--- the quantity which in Schwarzschild's method is stored in the orbit library.
In traditional M2M, one simply replaces $y_j$ with $\bar{y}_j$ directly in
equation (\ref{eq:foc:trad}). Even though, this means that weight adaption is
not strictly along the gradient of the merit function, it appears that the
method still converges in practice, though it is not completely obvious that it
always does (DL07 introduce the time-averaging only after they argue for
convergence), in particular for other forms of the constraint function than
simple $\chi^2$ expressions on model moments.

%%%%%%%%%%%%%%%%%%%%%%%%%%%%%%%%%%%%%%%%%%%%%%%%%%%%%%%%%%%%%%%%%%%%%%%%%%%%%%%%
\section{A novel M2M method}
\label{sec:novel}
In this section I criticise the traditional M2M method and propose alternatives
and/or modifications, which ultimately cumulate in a novel method.

\subsection{Time scales}
\label{sec:novel:time}
An important issue is the appropriate choice for the adjustment rate
$\epsilon$. The velocity of change (\ref{eq:foc}) varies on the orbital time
scale of the $i$th trajectory, because different parts of the orbit contribute
differently to the merit function. Clearly, the weight adaption should happen
adiabatically, i.e.\ $\epsilon <\Omega_i$ with $\Omega_i$ the natural
(orbital) frequency of trajectory $i$. Unfortunately, the orbital frequencies
of the $N$ trajectories may easily vary by many orders of magnitude, such that
meeting this condition for all of them becomes a serious problem. In
traditional M2M this is not really solved: using a very low adaption rate
$\epsilon$ ensures that the weights for all but the outermost trajectories are
adapted adiabatically.

One may think that using individual adjustment rates $\epsilon_i\propto\Omega_i$
would solve the problem. However, this is not the case: the method no longer
converges (it does initially, but eventually convergence stalls well before
reaching the optimum), presumably because such an alteration changes the
direction of adjustment away from the gradient of $Q$. Instead, I turn the
tables and achieve $\omega_i\propto\epsilon$ by integrating each trajectory on
its own dynamical time scale. To this end, I introduce the dimensionless time
\begin{equation}
  \tau = t/T_i
\end{equation}
with $T_i=2\pi/\Omega_i$ the orbital period, such that with respect to this new
dimensionless time each trajectory has natural frequency $\omega=2\pi$. The
equations of motion for the $N$-body system expressed in $\tau$ are
\begin{equation} \label{eq:eom:mod}
  \B{x}_i^{\prime\prime} = -T^2_i\,\B{\nabla} \Phi(\B{x}_i),
\end{equation}
where a prime denotes derivative w.r.t.\ $\tau$. Conversely, the M2M equation
remains:
\begin{equation} \label{eq:dotw:mod}
  w_i^\prime = \epsilon w_i U_i
\end{equation}
such that now $\epsilon$ is a dimensionless rate per orbit for \emph{each}
particle. In practice, a rough estimate for the orbital period, e.g.\ based on
the epicycle approximation, is sufficient for $T_i$.

\subsection{Enforcing total-weight conservation}
\label{sec:novel:mass}
With the traditional M2M formulation, conservation of the total weight is not
guaranteed, as the maximum of $Q$ may occur at $\sum_i w_i \neq 1$. This problem
has not been discussed by ST96 \& DL07, and I assume that it is dealt with by
simply re-normalising the weights after each step.

While this may be a viable method, I propose a somewhat different approach which
incorporates the total-weight constraint into the adjustment step. I start by
observing that the unconstrained maximum of the modified merit function
\begin{equation} \label{eq:Qstar}
  Q^\star(\B{w}) \equiv Q(\B{w}^\star) + \textstyle \ln \sum_kw_k - \sum_kw_k,
\end{equation}
maximises $Q(\B{w})$ subject to the constraint $1=\sum_k w_k$
\citep[e.g.][]{Dehnen1998}. Thus, the total-weight constraint can be
incorporated by replacing $Q$ with $Q^\star$ in equation (\ref{eq:foc}). Note
that since $Q^\star$ depends on the constraints only through the $w^\star_i$,
the $N$-body system must still be re-normalised after each adaption step, but
the step hardly carries the system away from normalisation.

Alternatively, the total weight of the $N$-body system may be allowed to float
freely and be constrained only by the data via the constraint function.

\subsection{An alternative adjustment}
\label{sec:novel:adjust}
As discussed in the last paragraph of \S\ref{sec:trad} the traditional
time-averaging procedure interferes with the computation of the gradient of
the merit function. Moreover, as I shall discuss in the next subsection, the
time-averaging is particularly undesirable when using the M2M method for
tailoring $N$-body initial conditions. These considerations lead me to
consider a different time-averaging approach: instead of averaging the
moments, I consider suppressing fluctuations in the merit function by
averaging $Q^\star$ (or $Q$) itself and its derivatives. In analogy to the
moment-averaging equation~(\ref{eq:mom:ave}), this would yield
\[
  U_i^\prime
  = \eta\left(\frac{\partial Q^\star}{\partial w_i} - U_i\right)
\]
with the dimensionless averaging rate $\eta$. Combining this with the
weight-adjustment equation (\ref{eq:dotw:mod}) results in a second-order
differential equation for $\varphi_i\equiv\ln w_i$:
\[
  \varphi_i^{\prime\prime} = \epsilon \eta
  \frac{\partial Q^\star}{\partial w_i} - \eta\,\varphi_i^\prime.
\]
Thus, $\partial Q^\star/\partial w_i$ acts like a force for the $\varphi_i$. I
now take this analogy even further and replace it with the gradient of
$Q^\star$ w.r.t.\ the dependent variable $\varphi_i$:
\begin{equation} \label{eq:ddphi:new}
  \varphi_i^{\prime\prime} = \lambda {\frac{\partial Q}{\partial\varphi_i}\!}^\star
  - \eta\,\varphi_i^\prime,
\end{equation}
(where I have substituted $\lambda$ for $\epsilon\eta$), corresponding to
\begin{equation} \label{eq:foc:alt}
  U_i^\prime = \eta\left(w_i\frac{\partial Q^\star}{\partial w_i} - U_i\right).
\end{equation}
For $\eta=0$, equation (\ref{eq:ddphi:new}) is equivalent to the familiar
equation of motion under the influence of the `potential' $-\lambda
Q^\star$. Thus, if the time-dependence of $Q^\star$ were solely due to
temporal changes in the weights, then the energy-like quantity
\begin{equation} \label{eq:Eps} \mathcal{E} \equiv \textstyle
  \frac{1}{2}\sum_i \varphi_i^{\prime2} - \lambda\,Q^\star 
\end{equation}
is conserved. The frictional term proportional to $\eta$ in equation
(\ref{eq:ddphi:new}) in fact ensures that $\mathcal{E}$ is not conserved but
decreases, thus ultimately leading to the maximum of $Q^\star$, as desired.
Thus, unlike the situation for traditionally M2M, where the time-averaging of
model moments may interfere with the convergence (see the discussion in the
last paragraph of \S\ref{sec:trad}), the time-averaging in my approach
provides the damping term required for convergence.

\subsection{\boldmath Tailoring $N$-body initial conditions}
\label{sec:novel:ics}
As already mentioned in the introduction, the M2M technique offers a natural
and powerful way to generate $N$-body initial conditions with prescribed
properties.  There is, however, a fundamental difference compared to employing
M2M for fitting data: the target values $Y_j$ now represent these prescribed
properties and, unlike observed data, have no natural uncertainties. A common
practice with model fits without known uncertainties is to simply set
$\sigma_j=1$ (alternatively, setting $\sigma_j=Y_j$ alters $C$ to measure the
relative error and yields ST96's original method). However, this is rather
unsatisfactory here, as the value obtained for $C$ then no longer provides
insight about the goodness of fit.

Moreover, unlike most parametric model fits, a $N$-body model, being a
Monte-Carlo representation, does have its own natural uncertainties. This
suggests that the $\sigma_j$ should be set to the uncertainties expected from
shot noise in the $N$-body model itself. If this is done, $C$ retains its
interpretability: a good fit corresponds to $C$ equalling to the number of
constraints. While this sounds natural and straightforward, it introduces some
subtle problems. One problem with traditional M2M is that the time-averaging
of the model moments intentionally reduces the shot noise, which invalidates
the interpretability of $C$.

With the alternative time-averaging of the merit function itself (see
\S\ref{sec:novel:adjust}), this is no longer the case, but the shot noise in the
$N$-body model causes temporal variations of $Q^\star$ additional to those
induced by changing the weights. This means that equation (\ref{eq:ddphi:new})
corresponds to following a frictional trajectory (in $N$-dimensional
$\B\varphi$-space) in a temporally fluctuating potential. The fluctuations are
of the same order as the optimal value for $C$ and prevent the algorithm to
converge in the sense that $\varphi_i^\prime\propto U_i\to0$. However, the
examples in \S4 suggest that this is not a serious problem.
% One may argue that this is not really a problem, since the properties of any
% $N$-body model are genuinely fluctuating.  However, one may want to suppress
% these fluctuations slightly by averaging the model moments similar to
% traditional M2M, though not in real time $t$ but in $\tau$, i.e.
% \begin{equation} \label{eq:ave:mod}
%   \bar{y}_j^\prime = \alpha (y_j - \bar{y}_j)
% \end{equation}
% with $\alpha$ a dimensionless averaging rate per orbital period.

\subsection{Re-sampling}
\label{sec:novel:res}
The adjustment of the weights in the M2M technique may lead to a wide range of
weights (or for $\hat{w}_i\neq N^{-1}$ to a wide range of
$w_i/\hat{w}_i$). This potentially reduces the effective resolution of the
$N$-body model substantially and is particularly undesirable if the $\sigma_j$
represent the uncertainties expected for a $N$-body system with weights
following the priors. A wide range in $w_i/\hat{w}_i$ (corresponding to
unequal masses for a flat prior) is also undesirable with $N$-body initial
conditions. Therefore, it is useful to re-sample the $N$-body model from time
to time during and after the adjustment process. This is easily done by
drawing phase-space points for the new model from the original set
$(\B{x},\B{\upsilon})_i$ with probability proportional to the relative
normalised weight
\begin{equation} \label{eq:phi}
  \gamma_i = w_{i,\mathrm{old}}^\star/\hat{w}_i,
\end{equation}
and subsequently setting the weights to $w_i=\hat{w}_i$ if the total weight is
constraint to unity and $w_i=\hat{w}_iN^{-1}\sum_kw_{k,\mathrm{old}}$ otherwise.

In this process, some trajectories of the original set will not be re-sampled,
others get copied exactly once, yet others several times. In this latter case, I
make the first copy a straight clone of the original phase-space point, but for
any additional copies, I first randomise position and velocity as far as the
underlying symmetry allows (for spherical symmetry, for instance, rotate them by
a random angle about a random axis), and secondly add a small random velocity
component. This added velocity component prevents multiple trajectories to be on
identical orbits, and allows the model to explore phase-space regions of high
weights.

In order for the M2M method to still maximise the same merit function, one has
to alter the definition of the pseudo-entropy to
\begin{equation} \label{eq:entropy:res}
  S = - \sum_i w_i^\star \log\frac{\Gamma_i\,w_i^\star}{\hat{w}_i}
\end{equation}
with $\Gamma_i$ the product of the factors $\gamma_i$ from each re-sampling so
far. In this way, the contributions to $S$ from each trajectory are on average
the same before and after re-sampling. However, the actual value for $S$ may
increase (in particular if a trajectory with $\gamma_i\ll1$ happens to be
re-sampled).

\subsection{Technicalities}
\label{sec:novel:tech}
The description of my M2M method, is completed by giving some technical
details.  The M2M adjustment step is taken to be $\delta\tau=2^{-6}$, which
appears to be sufficiently short. Between these, the trajectories are
integrated using individual adaptive time steps (which are required despite
the fact that every trajectory is integrated on its own orbital time). While
this could be implemented with any type of method, I use the traditional
$N$-body block-step scheme with a kick-drift-kick leap-frog and a time step
$\delta \tau = 2\pi f\sqrt{\B{x}^2/|\B{x}\cdot\B{\nabla}\Phi|}/T_i$ with
$f=1/400$. In this way, trajectories are automatically synchronised at M2M
adjustment steps and the simultaneous computation of gravitational forces for
many positions allows some optimisation.

The M2M equations are also integrated using a kick-drift-kick leap-frog---note
that equation (\ref{eq:foc:alt}) can be integrated exactly at fixed $w_i$. In
practice, $\epsilon$ is grown slowly over $\delta\tau=1$ to its final value,
but also limited to $\epsilon\le\eta\max\{|U_i|\}$ at any time.

The $N$-body model is re-sampled whenever the ratio between maximum and
minimum $w_i^\star/\hat{w}_i$ exceeds a certain threshold (4 in the runs of
\S\ref{sec:ics}) and a minimum interval has elapsed since the last re-sampling
($\delta\tau{\,=\,}10$ in \S\ref{sec:ics}). The phase-space coordinates for
the $k$th re-sampled trajectory are set to those of the $i$th original
trajectory where
\begin{equation}
  C_{i} < \bar{\gamma}(k-\tfrac{1}{2}) \le C_{i+1}, \qquad i,k\in[1,N]
\end{equation}
with $\bar\gamma\equiv N^{-1}\sum_k\gamma_k$ the mean relative normalised
weight and $C_i= \sum_{k<i}\gamma_k$ the cumulative relative normalised weight
of the original model. Trajectories with $\gamma_i<\bar\gamma$ generate at
most one copy, while those with $\gamma_i>\bar\gamma$ get copied once or
more. The random component added to the velocities of extra copies is drawn
from a normal distribution with standard deviation $0.05\exp(-\tau/10)$ times
the local escape velocity (but avoiding generation of unbound trajectories).
In the case of a flat prior $\hat{w}_i=N^{-1}$, which I used in
\S\ref{sec:ics}, these relations simplify somewhat (in particular
$\bar\gamma=N^{-1}$).

The M2M method is ideally suited for distributed-memory parallelisation, since
the gravitational potential is fixed (no interactions between trajectories)
and the evaluation of the merit function and its derivatives require only
minimal communications. I implemented my method using the message-passing
interface (\textsf{MPI}) and found the resulting code to be super-scaling:
doubling the number of processors at fixed problem size reduces the execution
time to less then half.  This is presumably a result of the increase in total
cache, reducing the total sum of computation times, which out-weighs the small
communication overhead.

%%%%%%%%%%%%%%%%%%%%%%%%%%%%%%%%%%%%%%%%%%%%%%%%%%%%%%%%%%%%%%%%%%%%%%%%%%%%%%%%

\section{\boldmath Application: tailored $N$-body initial conditions}
\label{sec:ics}
Almost all published $N$-body simulations featuring individual galaxies use
initially spherical dark-matter haloes with isotropic velocity distributions.
This is because for these settings distribution-function models, on which
$N$-body initial conditions are usually based, are relatively simple to obtain.
However, there is no physical justification for these simplifications and
triaxial dark haloes with anisotropic velocity distributions are certainly more
realistic. Here, I apply my novel M2M method to tailor such $N$-body initial
conditions.

%%%%%%%%%%%%%%%%%%%%%%%%%%%%%%%%%%%%%%%%%%%%%%%%%%%%%%%%%%%%%%%%%%%%%%%%%%%%%%%%
\begin{figure}
  \centerline{
    \resizebox{82mm}{!}{\includegraphics{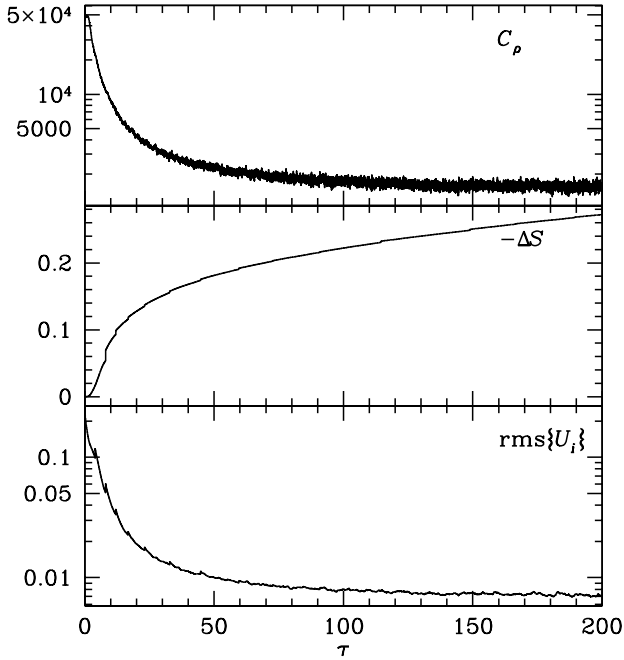}}
  }
  \caption{
    \label{fig:m2m:triaxial}
    Triaxial model: time evolution of the constraint function $C_\rho$,
    pseudo-entropy $S$, and rms value of the velocities of change for the M2M
    adjustment of $N=10^6$ particles with $\mu=100$, $\epsilon=0.5$,
    $\eta=0.5$.}
\end{figure} 
%%%%%%%%%%%%%%%%%%%%%%%%%%%%%%%%%%%%%%%%%%%%%%%%%%%%%%%%%%%%%%%%%%%%%%%%%%%%%%%%
\subsection{A triaxial halo model}
\label{sec:ics:triaxial}
Let us first consider the problem of designing a triaxial equilibrium with
prescribed shape and density profile, but without constraining its velocity
structure. The aim is to construct a triaxial truncated
\cite{DehnenMcLaughlin2005} model, which has density
\begin{equation} \label{eq:rho:halo}
  \rho \propto \left(\frac{q}{r_s}\right)^{-7/9}
  \left[\left(\frac{q}{r_s}\right)^{4/9}+1\right]^{-6}
  \mathrm{sech}\frac{q}{r_{\mathrm{t}}},
\end{equation}
with scale radius $r_s$, truncation radius $r_t$, and
`elliptical radius'
\begin{equation} \label{eq:ell:rad}
q^2\equiv \frac{x^2}{a^2} + \frac{y^2}{b^2} + \frac{z^2}{c^2}
\end{equation}
with $abc=1$. For this model, the radius at which
$-\mathrm{d}\ln\rho/\mathrm{d}\ln q=2$, often referred to as the scale radius
for dark-matter haloes, equals $r_2=(11/13)^{9/4}r_s\approx0.687r_s$. I choose
$r_t=10r_s$ and axis ratios $c/a=0.5$ and $b/a=0.7$.

%%%%%%%%%%%%%%%%%%%%%%%%%%%%%%%%%%%%%%%%%%%%%%%%%%%%%%%%%%%%%%%%%%%%%%%%%%%%%%%%
\begin{figure}
  \centerline{
    \resizebox{82mm}{!}{\includegraphics{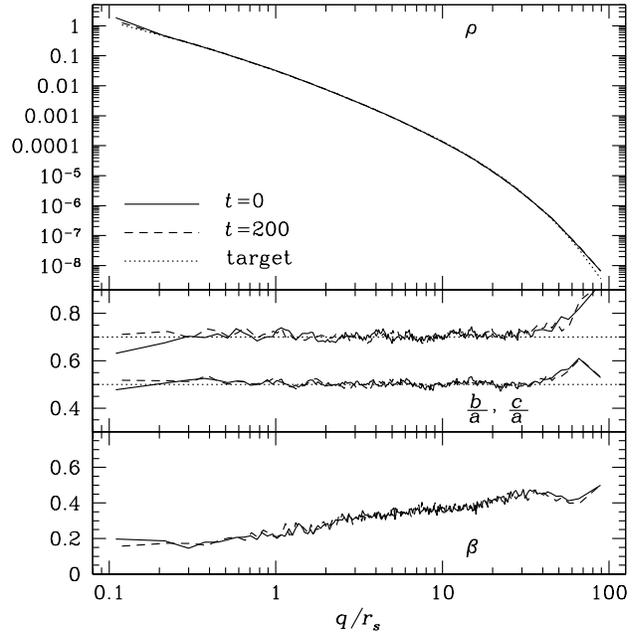}}
  }
  \caption{
    \label{fig:rho:triaxial}
    Triaxial model: density (top) and axis ratios (middle) plotted vs.
    elliptical radius for the target model (thin black lines) and the $N$-body
    model just after M2M adjustment ($t=0$) and after running them standalone
    (under self-gravity) for 200 time units. Also plotted is the velocity
    anisotropy parameter
    $\beta\equiv1-(\sigma_\theta^2+\sigma_\phi^2)/2\sigma_r^2$ (bottom).  Note
    that the density at very small and large values is overestimated.  an
    artefact of the density estimation procedure.}
\end{figure}
%%%%%%%%%%%%%%%%%%%%%%%%%%%%%%%%%%%%%%%%%%%%%%%%%%%%%%%%%%%%%%%%%%%%%%%%%%%%%%%%
A convenient way to constrain the full three-dimensional density distribution of
the model is by means of an expansion in bi-orthonormal potential-density basis
functions $\psi_{nlm}(\boldsymbol{x})$ and $\rho_{nlm}(\boldsymbol{x})$. These
satisfy the Poisson equation as well as the bi-orthonormality and completeness
conditions
\begin{eqnarray}
  \label{eq:pd:Poisson}
  -\boldsymbol{\nabla}^2\psi_{\boldsymbol{\mathsf{n}}}(\boldsymbol{x}) &=&
  4\pi\rho_{\boldsymbol{\mathsf{n}}}(\boldsymbol{x}), \\
  \label{eq:pd:orthonormal}
  \int \mathrm{d}^3\!\boldsymbol{x}\,
  \psi_{\boldsymbol{\mathsf{n}}}(\boldsymbol{x})\,
  \rho_{\boldsymbol{\mathsf{n}}^\prime}(\boldsymbol{x})
  &=& \delta_{\boldsymbol{\mathsf{n}}\boldsymbol{\mathsf{n}}^\prime}, \\
  \label{eq:pd:complete}
  \sum_{\boldsymbol{\mathsf{n}}}\psi_{\boldsymbol{\mathsf{n}}}(\boldsymbol{x})\,
  \rho_{\boldsymbol{\mathsf{n}}}(\boldsymbol{x}^\prime)
  &=& \delta(\boldsymbol{x}-\boldsymbol{x}^\prime),
\end{eqnarray}
where $\boldsymbol{\mathsf{n}}\equiv(n,l,m)$. In this study, I use
\citeauthor{Zhao1996}'s (\citeyear{Zhao1996}) basis set, whose lowest-order
functions satisfy
\begin{equation}
  \psi_{\boldsymbol{\mathsf{0}}}\propto\frac{1}{
    \left(|\boldsymbol{x}|^{1{\!/\!}a}+s^{1{\!/\!}a}\right)^{a}},\quad
  \rho_{\boldsymbol{\mathsf{0}}}\propto
  \frac{1}{|\boldsymbol{x}|^{2-1{\!/\!}a}
    \left(|\boldsymbol{x}|^{1{\!/\!}a}+s^{1{\!/\!}a}\right)^{2+a}}
\end{equation}
with scale radius $s$, and a free parameter $a$, which controls the density
profile. The expansion coefficients
\begin{equation}
  A_{\boldsymbol{\mathsf{n}}} = \sum_i w_i\,\psi_{\boldsymbol{\mathsf{n}}}(\boldsymbol{x}_i).
\end{equation}
are moments of the model and of the form (\ref{eq:y}), such that the resulting
constraint function
\begin{equation}\label{eq:C:rho}
  C_\rho = \sum_{\boldsymbol{\mathsf{n}}}
  \left(\frac{A_{\boldsymbol{\mathsf{n}}}-B_{\boldsymbol{\mathsf{n}}}}
    {\sigma_{\boldsymbol{\mathsf{n}}}}\right)^2
\end{equation}
is of the form (\ref{eq:C}). Note that the calculation of the derivative
\begin{equation}\label{eq:dC:rho}
  \frac{\partial C_\rho}{\partial w_i} = 2 \sum_{\boldsymbol{\mathsf{n}}}
  \frac{A_{\boldsymbol{\mathsf{n}}}-B_{\boldsymbol{\mathsf{n}}}}
    {\sigma_{\boldsymbol{\mathsf{n}}}^2} \psi_{\boldsymbol{\mathsf{n}}}(\boldsymbol{x}_i)
\end{equation}
is equivalent to computing the gravitational potential due to the coefficients
$2(A_{\boldsymbol{\mathsf{n}}}- B_{\boldsymbol{\mathsf{n}}}) /
\sigma_{\boldsymbol{\mathsf{n}}}^2$ at the position $\B{x}_i$. This has the
benefit that the functionality of an existing basis-function based $N$-body
force solver \citep[dubbed `self-consistent field code'
  by][]{HernquistOstriker1992} can be readily utilised. I use $s=r_s$ and
$a=9/4$ for the parameters of the expansion and include terms up to
$n_{\mathrm{max}}=20$ and $l_{\mathrm{max}}=12$. Since the model is forced to
have triaxial symmetry already by the assumed underlying gravitational
potential (see below), one does not need to constrain this symmetry.  This
considerably reduces the number of terms in equation (\ref{eq:C:rho}), since
coefficients with odd $l$ or $m$ can be ignored (they vanish for triaxial
symmetry) as well as those with $m<0$ (since $A_{n\,l\,m}=A_{n\,l\,-m}$),
which leaves just 588 independent constraints for $N=10^6$ particles.
%Furthermore, I choose to include into the constrain function only terms for
%which $|B_{\B{n}}|>\sigma_{\B{n}}$ (in general, this is not advisable, but in
%this particular case, I found no disadvantage compared to not imposing this
%restriction), which leaves just 170 independent constraints for $N=10^6$
%particles.

%%%%%%%%%%%%%%%%%%%%%%%%%%%%%%%%%%%%%%%%%%%%%%%%%%%%%%%%%%%%%%%%%%%%%%%%%%%%%%%%
\begin{figure}
  \centerline{ \resizebox{82mm}{!}{\includegraphics{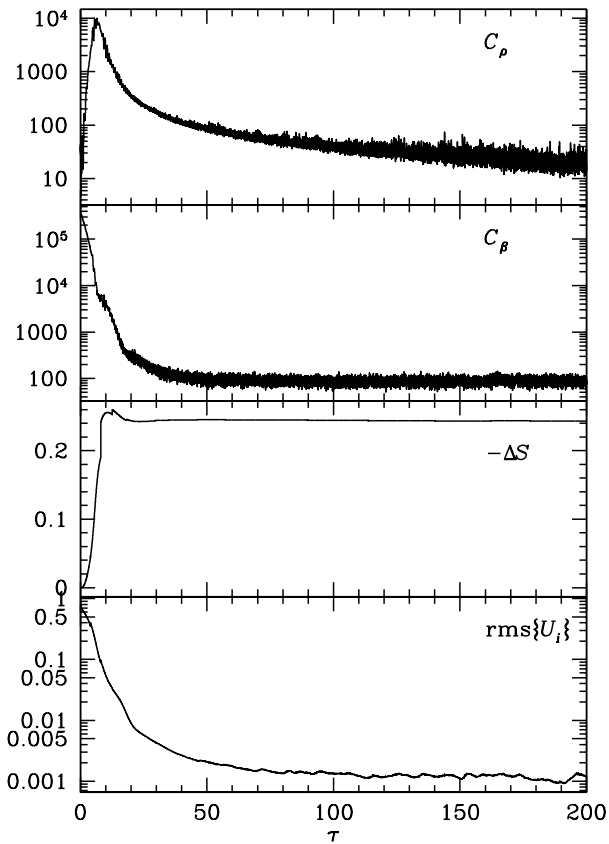}} }
  \caption{
    \label{fig:m2m:aniso}
    Similar to Fig.~\ref{fig:m2m:triaxial}, but for the M2M adjustment of the
    spherical model with anisotropic velocities.}
\end{figure} 
%%%%%%%%%%%%%%%%%%%%%%%%%%%%%%%%%%%%%%%%%%%%%%%%%%%%%%%%%%%%%%%%%%%%%%%%%%%%%%%%
In order to prepare for the M2M adjustment, I sample $N=10^6$ positions from
the target density model and evaluate the resulting $A_{\boldsymbol{\mathsf{n}}}$. This
is repeated many times to obtain the target values $B_{\boldsymbol{\mathsf{n}}}$ and
the expected errors $\sigma_{\boldsymbol{\mathsf{n}}}$ from ensemble averaging. Next,
also the velocities are sampled from the equivalent spherical (and
velocity-isotropic) distribution-function model and scaled in each dimension
such that the tensor virial theorem is satisfied (whereby correcting for
unbound particles). Lastly, to achieve phase-mixing the resulting trajectories
are integrated for several orbital times in the potential of the target model
computed from the expansion
\begin{equation} \label{eq:pot}
  \Phi(\B{x}) = -G M_{\mathrm{tot}}\sum_{\boldsymbol{\mathsf{n}}} B_{\boldsymbol{\mathsf{n}}}\,
  \psi_{\boldsymbol{\mathsf{n}}}(\boldsymbol{x}).
\end{equation}
I hoped this procedure already results in a model close to the target, but this
was not the case at all: the resulting value for the constraint function
(\ref{eq:C:rho}) is $\sim5\times10^4\gg588$.

Finally, I run the M2M scheme of equations (\ref{eq:dotw}) and
(\ref{eq:foc:alt}) with $\epsilon=0.5$, $\eta=0.5$,
%with or without moment averaging (equation \ref{eq:ave:mod}),
and various values for $\mu$, whereby integrating the trajectories in the target
potential (\ref{eq:pot}). Fig.~\ref{fig:m2m:triaxial} shows the time evolution
of $C_\rho,\,S,$ and the rms value of $U_i$ for an adjustment run with
$\mu=100$. After a quick reduction of the error (as measured by $C_\rho$),
convergence becomes somewhat slower. $C_\rho$ fluctuates with amplitude similar
to its good-fit value of 588, as the discussion in \S\ref{sec:novel:ics}
suggested, and does not reach this value. 

In order to independently assess whether the adjustment successfully produced
a stable $N$-body model with the desired properties, I run it for 200 time
units\footnote{I use a unit system with $r_s=1$, $G=1$, and $M=1$, the total
  halo mass.} (corresponding to $\sim6$ dynamical times at the scale radius)
under self-gravity, whereby monitoring shape and density profile. This latter
is done by first estimating the density at each particle using a
kernel-estimator from its 32 nearest neighbours and then binning particles in
density (with 5000 per bin) to estimate the rms radius and axis ratios (from
the eigenvalues of the moment-of-inertia tensor) of density shells. As evident
from Fig.~\ref{fig:rho:triaxial}, the $N$-body model matches the target very
well, except for too little flattening in the outermost regions. It appears
thus, that the failure of convergence of $C_\rho$ close to its good-fit value
is caused by the model being too round at very large radii (well outside the
virial radius of a CDM halo). The model appears stable: there is no
significant change over 200 time units.

\begin{figure}
  \centerline{
    \resizebox{82mm}{!}{\includegraphics{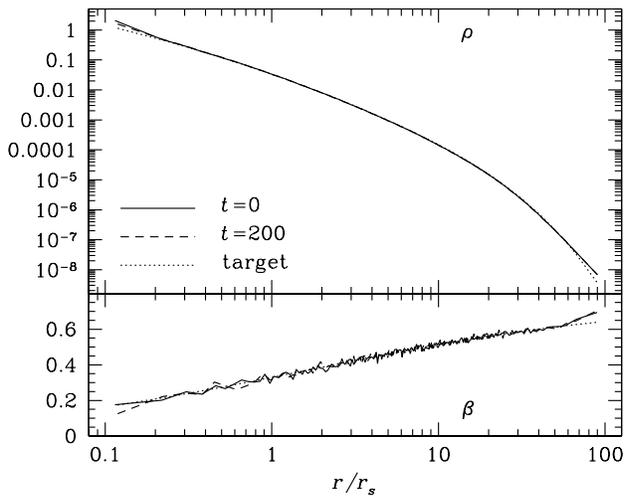}}
  }
  \caption{
    \label{fig:rho:aniso}
    Spherical anisotropic model: radial profiles of density and velocity
    anisotropy parameter for the target model (thin black lines) and those
    measured for the $N$-body model just after M2M adjustment ($t=0$) and
    after running them standalone (under self-gravity) for 200 time units.}
\end{figure} 
%%%%%%%%%%%%%%%%%%%%%%%%%%%%%%%%%%%%%%%%%%%%%%%%%%%%%%%%%%%%%%%%%%%%%%%%%%%%%%%%
%%%%%%%%%%%%%%%%%%%%%%%%%%%%%%%%%%%%%%%%%%%%%%%%%%%%%%%%%%%%%%%%%%%%%%%%%%%%%%%%
\begin{figure}
  \centerline{
    \resizebox{82mm}{!}{\includegraphics{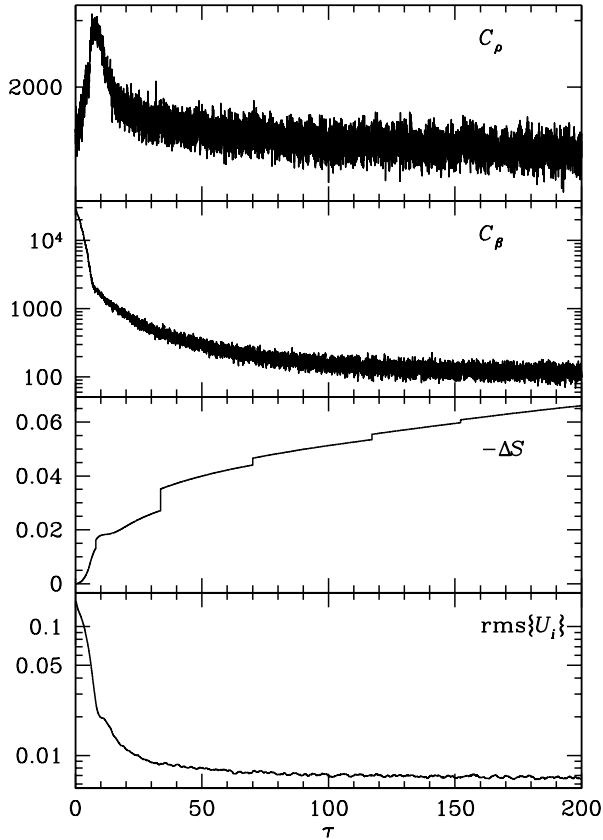}}
  }
  \caption{
    \label{fig:m2m:triaxial:aniso}
    Similar to Figs.~\ref{fig:m2m:triaxial} and \ref{fig:m2m:aniso}, but for
    M2M adjustment of triaxial model with anisotropic velocities.}
\end{figure} 
%%%%%%%%%%%%%%%%%%%%%%%%%%%%%%%%%%%%%%%%%%%%%%%%%%%%%%%%%%%%%%%%%%%%%%%%%%%%%%%%

\subsection{A spherical halo model with anisotropic velocities}
\label{sec:ics:aniso}
Next I consider the problem of generating a spherical halo model with
specified velocity anisotropy. I use the spherical version of the model
(\ref{eq:rho:halo}) and aim to constrain Binney's anisotropy parameter
$\beta\equiv1-(\sigma_\theta^2+\sigma_\phi^2)/2\sigma_r^2$ to have radial
profile
\begin{equation} \label{eq:beta:model}
  \beta_{\mathrm{model}}(r) =
  \beta_\infty \frac{(r/r_s)^{4/9}}{1 + (r/r_s)^{4/9}}.
\end{equation}
This corresponds to isotropy ($\beta=0$) in the very centre, and a slowly
increasing radial anisotropy (for $\beta_\infty>0$), reaching
$\beta\to\beta_\infty$ at $r\to\infty$, which describes simulated dark-matter
haloes remarkably well \citep{DehnenMcLaughlin2005}. If one wants to retain
the traditional M2M approach of constraining model moments, one must constrain
the moments $\rho\sigma^2_r$, $\rho\sigma^2_\theta$, and $\rho\sigma^2_\phi$
to values obtained from solving the Jeans equation. However, this latter step
requires spherical symmetry and hence cannot be generalised to non-spherical
systems.  Instead, I directly constrain the anisotropy via
\begin{equation}
  C_\beta = \sum_j
  \left(\frac{\beta_j-\beta_{\mathrm{model}}(r_j)}
    {\sigma_j^2}\right)^2,
\end{equation}
where $j$ indexes radial bins with rms radius $r_j$ and measured anisotropy
\begin{equation}
  \beta_j = 1 - \frac
  {\sum_i w_i (\upsilon_{\theta i}^2+\upsilon_{\phi i}^2)}
  {2 \sum_i w_i \upsilon_{r i}^2},
\end{equation}
where the sums are over all particles in the $j$th radial bin. Since $\beta_j$
is not a moment of the model, but a function (ratio) of moments, it is not of
the form (\ref{eq:y}). However, the derivatives $\partial \beta_j/\partial
w_i$, needed for $\partial C_\beta/\partial w_i$, can still be easily
computed. The uncertainties $\sigma_j$ could be estimated from the $N$-body
model, but since such an estimate depends on the $w_i$ and hence adds to
$\partial C_\beta/\partial w_i$, this would complicate matters
unnecessarily. Instead, I simply assume (with $n_j$ the number of particles in
the bin)
\begin{equation}\label{eq:sigma:beta}
  \sigma_j = (1-\beta)\sqrt{\frac{3}{n_j-1}},
\end{equation}
which is the standard deviation expected for a multivariate normal velocity
distribution\footnote{This standard deviation also depends on the ratio
  $\sigma_\theta^2/\sigma_\phi^2$. The minimum occurs for
  $\sigma_\theta^2=\sigma_\phi^2$ and corresponds to
  equation~(\ref{eq:sigma:beta}), while the maximum (arising at vanishing
  $\sigma_\theta^2$ or $\sigma_\phi^2$) is only a factor $\sqrt{4/3}$ larger.}
with anisotropy $\beta$.

Fig.~\ref{fig:m2m:aniso} shows the time evolution of the various quantities
for an M2M run with $N=10^6$, $n_{\mathrm{max}}=20$ in $C_\rho$ (only terms
with $l=m=0$ are considered) and 100 radial bins for $C_\beta$ with
$\beta_\infty=0.75$, i.e.  increasing radial anisotropy. The values for the
constraint functions converged to their best-fit values relatively quickly and
after $\tau\sim100$ hardly any improvement is made. This is different from the
situation for the triaxial halo model in the previous sub-section, which took
much longer to reduce $C_\rho$ and required much smaller $\mu$. The reason for
this difference is not clear, but possibly it is because the solution space
for spherical models with anisotropic velocities is larger than that for
triaxial models with the assumed axis ratios.

The resulting radial profiles for $\rho$ and $\beta$ for the $N$-body model
provide an excellent match to the target values both just after the M2M
adjustment and after running the model in isolation (under self-gravity) for
200 time units, as demonstrated in Fig.~\ref{fig:rho:aniso}.

%%%%%%%%%%%%%%%%%%%%%%%%%%%%%%%%%%%%%%%%%%%%%%%%%%%%%%%%%%%%%%%%%%%%%%%%%%%%%%%%
\begin{figure}
  \centerline{
    \resizebox{82mm}{!}{\includegraphics{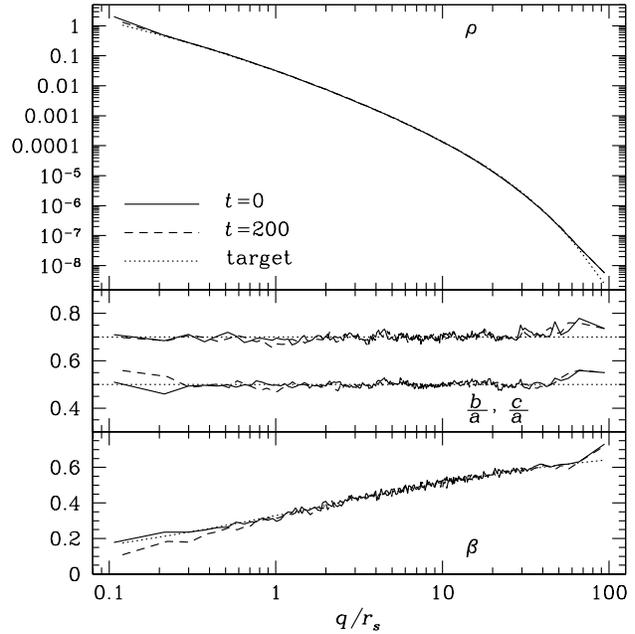}}
  }
  \caption{
    \label{fig:rho:triaxial:aniso}
    Triaxial model with anisotropic velocities: radial profiles of density,
    axis ratios, and velocity anisotropy parameter for the target model (thin
    black lines) and those measured for the $N$-body model just after M2M
    adjustment ($t=0$) and after running them standalone for 200 time units.}
\end{figure} 
%%%%%%%%%%%%%%%%%%%%%%%%%%%%%%%%%%%%%%%%%%%%%%%%%%%%%%%%%%%%%%%%%%%%%%%%%%%%%%%%

\subsection{A triaxial halo model with anisotropic velocities}
\label{sec:ics:triaxial:aniso}
Finally, I want to generate a triaxial halo with the same triaxial density
distribution as in \S\ref{sec:ics:triaxial} but with the velocity anisotropy
profile given by equation (\ref{eq:beta:model}) with $\beta_\infty=0.75$,
though with $r$ replaced with the elliptical radius $q$ defined in equation
(\ref{eq:ell:rad}). To this end, I take the $N$-body models generated in
\S\ref{sec:ics:triaxial} as starting point for the M2M adjustment. Just as in
the previous sub-section, the constraint function now consists of two terms,
constraining the density and velocity anisotropy, respectively.
Figs.~\ref{fig:m2m:triaxial:aniso} and \ref{fig:rho:triaxial:aniso} show,
respectively, the M2M adjustment and the comparison of the final model with
the target. Evidently, the model matches the target excellently, even the
shape in the outermost parts, which was too round in \S\ref{sec:ics:triaxial}.

%%%%%%%%%%%%%%%%%%%%%%%%%%%%%%%%%%%%%%%%%%%%%%%%%%%%%%%%%%%%%%%%%%%%%%%%%%%%%%%%
\section{Discussion}
\label{sec:disc}
The basic idea of the made-to-measure (M2M) method is to adjust the $N$-body
weights until the model satisfies some constraints, expressed as maximisation
of a merit function. This is achieved by changing the weights slowly in the
uphill direction of the merit function. While any M2M algorithm must follow
this basic recipe, there is significant freedom in the details of how this is
done. The purpose of this study was to improve these details compared to the
original method as proposed by \cite{SyerTremaine1996} and slightly further
developed by \cite{DeLorenziEtal2007}.

A significant problem of this original method originates from the fact that
the natural time scale for the adjustment (and moment-averaging) is some small
multiple of the orbital time. Since the latter varies substantially between
orbits and, in particular, has no finite upper limit, any finite value for the
adjustment time results in too slow or too fast an adjustment for most
orbits. I solved this problem by introducing a dimensionless time variable,
which effects to integrating each trajectory for the same number of orbital
times. This is similar to Schwarzschild's method, where usually each orbit
considered is integrated for a fixed amount of orbital times. In fact, the
number of orbital time scales in Schwarzschild's method is of the same order
($\sim100$, depending on details, such as orbital symmetries) as with my M2M
technique, indicating that for typical orbits this number is required to
gather sufficient information.

Note that the iterative method of \cite{RodionovAthanassoulaSotnikova2009},
mentioned in the introduction, suffers from the same basic problem: evolving
the model over some (short) time scale will bring the inner parts, where the
dynamical time is short, much closer to equilibrim than the outer parts. In
order to gather sufficient information about the dynamics in the outer parts,
one would need to integrate orbits (or evolve the model) over a considerably
longer time than is often practical.

Another issue with the original M2M method is that the averaging of the model
moments, required to suppress $N$-body shot noise, interferes with the
adjustment process, though apparently this did not lead to practical problems
so far. However, if the M2M method is used to tailor $N$-body initial
conditions, the uncertainties entering the constraints are not observational
errors but those due to shot noise in the $N$-body model itself. In this case,
time-averaging the model moments reduces this shot noise and renders the
interpretation of the $\chi^2$-like constraint meaningless. I have overcome
both these problems by introducing a novel adjustment algorithm which effects
to time-averaging the merit function instead of the model moments and
corresponds to following an orbit in $N$-dimensional weight space with the
merit function representing the potential. A damping term, which emerges from
the averaging, guarantees that the maximum of the merit function will be
reached.

I also propose to (optionally) modify the merit function such that it
automatically meets the total-mass conservation constraint. Finally, I propose
to re-sample the $N$-body model from time to time during the M2M adjustment
process to (1) avoid a loss of resolution because of unequal weights, and (2)
to allow the model to explore phase-space in regions of high weights. This
latter is achieved by adding a small random velocity to extra copies of
trajectories, effecting to probe another orbit close to a highly weighted one.

Certainly, one can think of further improvements to the M2M method. One issue
is an automatic adaption of the parameters $\epsilon$ and $\eta$ to achieve
optimal convergence (a technique to adapt $\mu$ such that the constraint
function obtains a certain numerical value was already proposed by ST96). The
priors for the weights can be used to allow the $N$-body model to have
different mass resolution in different phase-space regions, which is a common
technique \citep[e.g.][]{ZempEtAl2008,ZhangMagorrian2008} for increasing the
resolution in, say, the inner parts of models for dark-matter haloes. A
significant speed-up may be achieved by starting with a relatively low number
of particles and increasing $N$ (essentially like re-sampling) only later
after the merit function is close to maximal.

Finally, one would like to adapt not only the weights of the $N$-body model,
but also the underlying mass distribution generating the gravitational
potential, for instance when interpreting kinematic data in terms of the
underlying (dark-) matter distribution. Unfortunately, changes in the orbits
induced by changes of the gravitational potential are not straightforward to
anticipate and hence to take into account in the adjustment process. In a
spherical setting one may, for instance, re-scale the phase-space coordinates
of every particle such that the eccentricity, inclination, and mean radius is
preserved when the gravitational potential is changed. Unfortunately, however,
something similar can no longer be done in the general, i.e.\ triaxial,
case. Thus, it seems that this is a really hard problem and that one is forced
to `jump' from one mass model to the next whereby starting from the best-fit
$N$-body model of a `nearby' potential. In this case, convergence may be fast,
i.e.\ only a few ten orbital times, leading to significant speed-up.

As the practical examples of the previous section demonstrated, my novel M2M
algorithm is a powerful tool to construct $N$-body models with specified
properties. One may use the method to explore possible stellar-dynamical
equilibrium solutions and their properties. For instance, the triaxial models
of \S\ref{sec:ics:triaxial} exhibit a significant radial velocity anisotropy
(Fig.~\ref{fig:rho:triaxial} bottom panel), even though the initial conditions
fed to the M2M adjustment procedure were created from a velocity isotropic
model and the velocity structure was not constrained. This strongly suggests
that radial velocity anisotropy is an inevitable property of (non-rotating)
triaxial equilibrium models. This can be qualitatively understood from the
inevitable prevalence of box orbits, which are the only orbital family
supporting a triaxial shape, but a more quantitative understanding would be
desirable.

%%%%%%%%%%%%%%%%%%%%%%%%%%%%%%%%%%%%%%%%%%%%%%%%%%%%%%%%%%%%%%%%%%%%%%%%%%%%%%%%
\subsection*{Acknowledgements}
Research in theoretical astrophysics at Leicester is supported by a STFC
rolling grant.
%%%%%%%%%%%%%%%%%%%%%%%%%%%%%%%%%%%%%%%%%%%%%%%%%%%%%%%%%%%%%%%%%%%%%%%%%%%%%%%%
%\bibliographystyle{mn2e}
%\bibliography{refs}
%\label{lastpage}
%\end{document}

\label{lastpage}
\end{document}